\documentclass[12pt]{article}

\usepackage{graphicx}

\newcommand\pubnumber{DPF2015-207}

\newcommand\pubdate{\today}

\def\louisville{Department of Physics \&\ Astronomy\\  University of Louisville, Louisville, KY  40292  USA}

\def\support{\footnote{Work supported by the Office of Science, U.S. Department of Energy, under contract DE-FG02-13ER41932}}

\usepackage{relsize}

\def\babar{\mbox{\slshape B\kern-0.1em{\smaller A}\kern-0.1em

    B\kern-0.1em{\smaller A\kern-0.2em R}}}

\def\Title#1{\begin{center} {\Large #1 } \end{center}}

\def\Author#1{\begin{center}{ \sc #1} \end{center}}

\def\Address#1{\begin{center}{ \it #1} \end{center}}

\newcommand\pubblock{\rightline{\begin{tabular}{l} \pubnumber\\        \pubdate  \end{tabular}}}

\newenvironment{Abstract}{\begin{quotation}  }{\end{quotation}}

\newenvironment{Presented}{\begin{quotation} \begin{center} 
             PRESENTED AT\end{center}\bigskip 
      \begin{center}\begin{large}}{\end{large}\end{center} \end{quotation}}

\def\Acknowledgments{\bigskip  \bigskip \begin{center} \begin{large}
             \bf ACKNOWLEDGMENTS \end{large}\end{center}}

\def\beq{\begin{equation}}
\def\eeq#1{\label{#1}\end{equation}}
\def\eeqn{\end{equation}}

\def\beqa{\begin{eqnarray}}
\def\eeqa#1{\label{#1}\end{eqnarray}}
\def\eeqan{\end{eqnarray}}

\let\bar=\overbar

\def\Dslash{\not{\hbox{\kern-4pt $D$}}}
\def\dslash{\not{\hbox{\kern-2pt $\del$}}}

\def\msb{{\bar{\ssstyle M \kern -1pt S}}}

\begin{document}

\begin{titlepage}

\pubblock

\vfill
\Title{Search for a dark photon at {\it \babar}}
\vfill

\Author{ David Norvil Brown\support \\  On behalf of the \babar\ Collaboration}
\Address{\louisville}
\vfill

\begin{Abstract}
\noindent We present the results of a search for a dark photon
decaying to $e^+e^-$ or $\mu^+\mu^-$ in the \babar\ detector~\cite{ORIGINAL}.  We find
no evidence for such a dark photon and set upper limits on the dark mixing as a function mass, tightening the constraints on the parameter 
space of several dark sector theories.

\end{Abstract}

\vfill

\begin{Presented}
DPF 2015\\
The Meeting of the American Physical Society\\
Division of Particles and Fields\\
Ann Arbor, Michigan, August 4--8, 2015\\
\end{Presented}

\vfill

\end{titlepage}

\def\thefootnote{\fnsymbol{footnote}}

\setcounter{footnote}{0}

\section{Introduction}

The notion of dark matter was introduced over 80 years ago by Zwicky~\cite{Zwicky}.
In the intervening years, the astrophysical evidence for dark matter has 
mounted, but the precise nature of this matter is still not understood.  
In the 1980's, the idea of a new force corresponding to a new $U(1)'$ symmetry was introduced~\cite{FifthForce}.  This idea was adapted to the dark sector, suggesting a dark gauge boson.
Recent astronomical observations, particularly of an excess of electrons
from cosmic rays~\cite{ASTRO}, has raised interest in the possibility of a dark gauge boson that may couple via kinetic mixing to Standard Model (SM) 
particles and could have a mass in the few ${\rm GeV/c^2}$ range.  The gauge boson is generally referred to as a dark photon ($A'$)
and can be considered to appear in any process in which a SM photon could appear.  
The coupling constant for such a dark photon vertex is taken to be $\alpha' = \epsilon^2\alpha$.
Current opinion generally places $\epsilon$ in the range 
$10^{-2} - 10^{-7}$.

\section{The \babar\ detector and data sets}

The \babar\ detector is a general-purpose collider detector consisting of five sub-detectors:  the 
Silicon Vertex Tracker (SVT); the Drift Chamber (DCH); the Detector for Internally Reflected Cherenkov radiation (DIRC); the Electromagnetic Calorimeter (EMC); and the Instrumented Flux Return (IFR).   The SVT, DCH, DIRC, and EMC reside inside a superconducting solenoid that provides a uniform 1.5-Tesla axial
magnetic field.  The SVT and DCH combine to provide tracking for charged particles and particle identification through specific ionization measurements.  The SVT also provides excellent vertex resolution
while the 40-layer DCH allows measurement of the curvature of tracks in the magnetic field, providing 
transverse momentum determination.  The DIRC is used for charged particle identification, especially 
for separating pions and kaons, and the EMC detects positions and energies of photons and electrons.
A detailed description of the \babar\ detector can be found in~\cite{NIMA}.

The data sample used in this analysis corresponds to an integrated luminosity of $514 {\rm fb^{-1}}$ 
recorded by the \babar\ experiment using the PEP-II $e^+e^-$ storage rings at the SLAC National Accelerator Laboratory.  
This data set was recorded between 1999 and 2008 and contains data primarily from the $\Upsilon(4S)$ 
as well as the $\Upsilon(3S)$ and $\Upsilon(2S)$, and data taken off-peak.  Off-peak data was taken 40 MeV below
the $\Upsilon(4S)$ and 30 MeV below the $\Upsilon(3S)$ and $\Upsilon(2S)$.

\section{Analysis Procedure}

In this analysis, we search for the process $e^+e^- \to\ \gamma A'$ with subsequent decay 
$A' \to\ e^+e^-,\ \mu^+\mu^-$.  The topology of this event is relatively simple and experimentally clean.  We search for events with one photon and 
two oppositely charged leptons.  %
We search for dark photon candidates in
the mass range $0.02\ {\rm GeV/c^2} - 10.2\ {\rm GeV/c^2}$.

The signal and background efficiencies are studied using a variety of simulations.  Signal events are simulated with MadGraph~\cite{MADGRAPH} for 
35 different hypothetical $A'$ masses.  Primary non-resonant backgrounds to the 
electron modes are simulated with BHWIDE~\cite{BHWIDE} while those
for muon modes are simulated with KK2f~\cite{KK}.   Simulation of the decays from resonances such as $\Upsilon(nS)$, $j/\psi$, etc.~formed in initial state radiation is
handled using structure function techniques~\cite{RESON}.
A GEANT4-based~\cite{G4} simulation is used to model material interactions and detector
 response.  A subset of approximately 5\%\ of the data sets is used to tune selection criteria and study systematic effects.  This data is then discarded and not used in the final 
 results.

The photon used for event selection must have an energy greater than 200 MeV.  We allow extra photons as long as their 
energies do not add to more than 200 MeV.
We consider only events in which the invariant mass of the photon and charged particles is consistent with the beam energy spread and in which 
the interaction point of the particles are consistent with the beam spot location.  We accept only those events in which the candidate kinematic vertex fit 
has a $\chi^2 < 30$.  We require at least one charged particle be well-identified as an 
electron and one loosely consistent with the electron hypothesis, or both tracks to be 
well-identified as muons.  

A significant background comes from radiative Bhabhas -- events of the type $e^+e^- \to\ e^+e^-\gamma$ -- in which one of the final state leptons has 
radiated the 
photon.  To suppress this background, we additionally require that the final state electron and positron form an angle with
their respective incoming beams such that the cosine of this angle is less than 0.5.  

A sizable background from converted photons in $e^+e^- \to\ \gamma\gamma$, $\gamma \to\ e^+e^-$ events still remains for low $e^+e^-$ invariant 
mass.  A neural network is employed to further reduce this background using five kinematic variables.   The selected cut on the neural net output 
keeps over 70\%\ of remaining signal while removing 99.7\%\ of the conversion background in the $m_{A'} \approx\ 20-50\ {\rm MeV/c^2}$ range.

\begin{figure}[htb]
\centering
\includegraphics[width=5.5in]{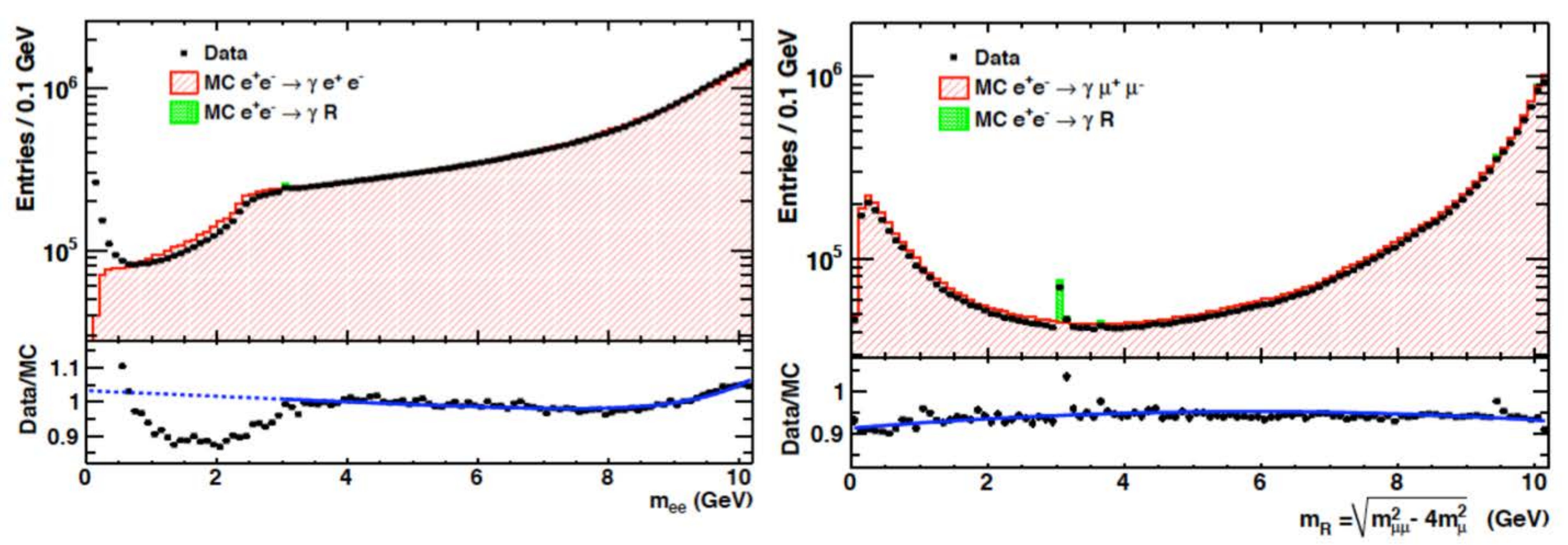}
\caption{Distribution of the final dielectron (left) and reduced dimuon invariant masses (right) as black points, overlaid with the predictions
of various SM processes (red) and ISR production of resonances (green).}
\label{fig:specs}
\end{figure}

The dielectron and reduced dimuon mass distributions are displayed in Figure~\ref{fig:specs}, overlaid with the predictions of various simulated
SM processes, including several resonances.  The reduced dimuon mass, $m_R = \sqrt{m_{\mu\mu}^2 - 4 m_{\mu}^2}$, is used in place of the 
invariant mass
because it is easier to model near threshold.  The data are seen to match well to the predictions from simulation, except in the very low $e^+e^-$ region, where BHWIDE is known to have difficulties.

Signal efficiency is determined from Monte Carlo simulation and found to be approximately 15\%\ for the dielectron channel and 35\%\ for the dimuon channel.  The difference is mainly attributable to trigger efficiencies, where a trigger filter reduces the rate of radiative Bhabha events.

Signal yield as a function of dark photon mass is extracted by performing a series of independent fits to the mass spectra for each beam energy.
The fits are performed in the range $0.02\ {\rm GeV/c^2} < m_{A'} < 10.2\ {\rm GeV/c^2}$ for the dielectron sample and $0.212\ {\rm GeV/c^2} < m_{A'} < 10.2\ {\rm GeV/c^2}$ for the dimuon sample.  We search for the dark photon in steps of approximately half the dark photon mass resolution with each fit window typically 30 times broader than the the signal resolution.  The signal resolution is estimated from a range of simulated
$A'$ samples and varies from 1.5 ${\rm MeV/c^2}$ to 8 ${\rm MeV/c^2}$.  A total of 5704 fits are performed in the dielectron channel and 5370 are performed in the
dimuon channel.  The fit bias is estimated from a large collection of toy-Monte Carlo pseudo-experiments and found to be negligible.

\begin{figure}[htb]
\centering
\includegraphics[width=5.5in]{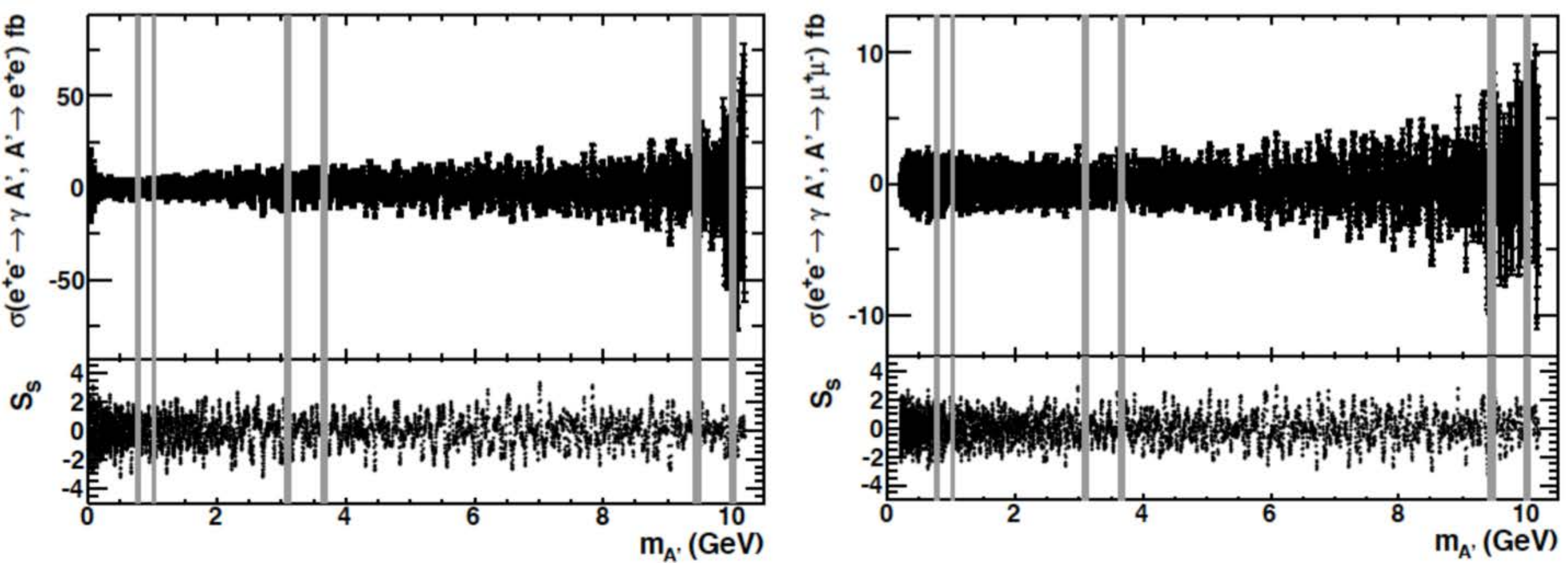}
\caption{Fitted cross-sections for dielectron (top) and dimuon (bottom) modes, together with their statistical significances, as functions
of the dark photon mass.  Grey bands indicate
mass regions that are excluded from the analysis.}
\label{fig:cross}
\end{figure}

\begin{figure}[htb]
\centering
\includegraphics[width=4.5in]{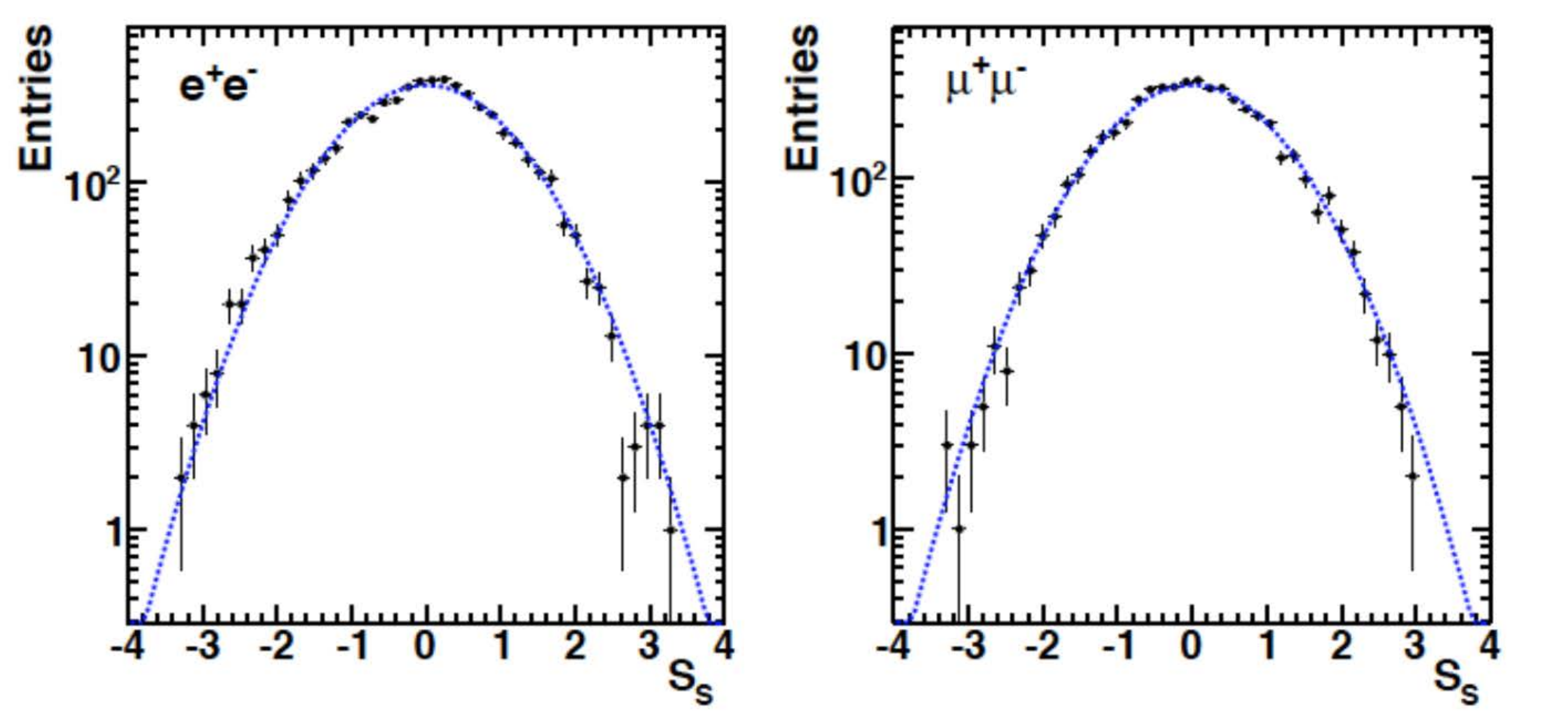}
\caption{Distribution of the statistical significance, $S$ from the fits to the dielectron (left) and dimuon (right) final states, together with the 
expected distribution for the null hypothesis (dotted line).}
\label{fig:sigs}
\end{figure}

A statistical significance is assigned to each fit as $S = \sqrt{2\log({\mathcal L}/{\mathcal L_0})}$, where ${\mathcal L}$ is the likelihood for the fit 
with background plus signal and ${\mathcal L_0}$ is the likelihood with a background function only.  The statistical significance is assigned a sign 
that is the same as the sign of the area of the fitted signal function.  The largest local significances are $3.4\sigma$
($2.9\sigma$) near $m_{A'} = 7.02\ {\rm GeV/c^2}\ (6.09\ {\rm GeV/c^2})$ for the dielectron (dimuon) final state.  Including trial factors, studied via
a large number of toy Monte Carlo experiments, we find the corresponding p-value is 0.57 (0.94), consistent with the null hypothesis.
We extract the cross-section as a function of dark photon mass by dividing the fitted yields by efficiency and luminosity.  The cross-sections are shown
in Figure~\ref{fig:cross} and the distributions of statistical significance are shown in Figure~\ref{fig:sigs}.

\begin{figure}[htb]
\centering
\includegraphics[width=3.5in]{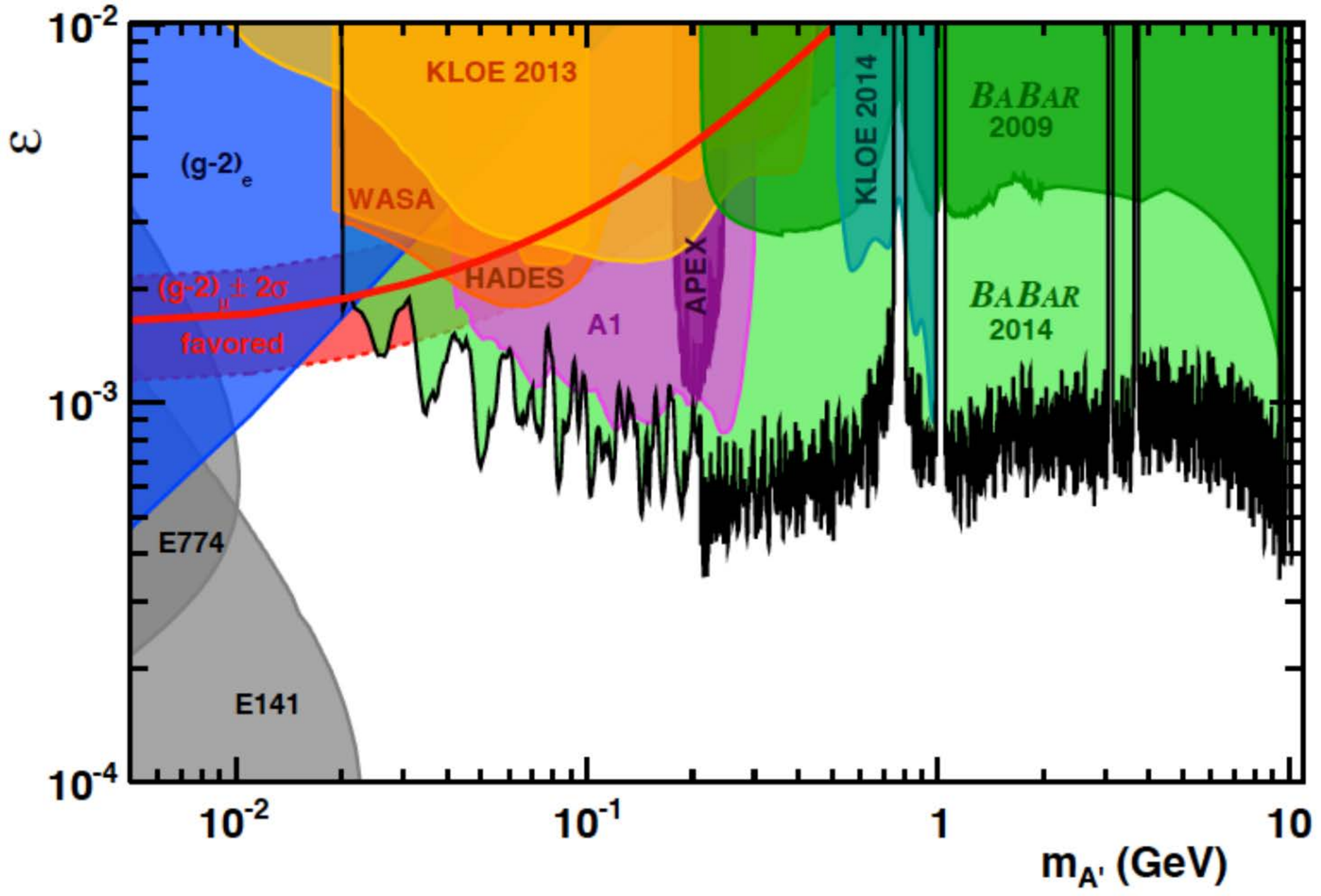}
\caption{Upper limits at 90\%\ confidence level on the kinetic mixing strength, $\epsilon$, as a function of dark photon mass.}
\label{fig:paramspace}
\end{figure}

Using the expected dark photon branching fractions for $A' \to\ l^+l^-$~\cite{FRAX}, our results can be combined and translated  
into a 90\%\ confidence level upper limits on the kinetic mixing strength.  %
These results~\cite{ESSIG} are 
shown in Figure~\ref{fig:paramspace}, which illustrates clearly that our measurements significantly increased the constraints on the 
dark sector parameter space.

\Acknowledgments
The \babar\ Collaboration would like to thank Rouven Essig, Sarah Andreas, Philip Schuster, and Natalia Toro for 
information essential to this work, and for access to the MadGraph code for simulaiton.
We are grateful for the excellent luminosity
and machine conditions provided by our PEP-II
colleagues, and for the substantial dedicated effort from
the computing organizations that support \babar.  The
collaborating institutions wish to thank SLAC for its
support and kind hospitality.
This work is supported
by DOE and NSF (USA), NSERC (Canada), CEA and
CNRS-IN2P3 (France), BMBF and DFG (Germany),
INFN (Italy), FOM (The Netherlands), NFR (Norway),
MES (Russia), MICIIN (Spain), STFC (United Kingdom).
Individuals have received support from the Marie
Curie EIF (European Union), the A.~P.~Sloan Foundation
(USA) and the Binational Science Foundation (USA--Israel).

\end{document}